\def\kms{\relax \ifmmode {\,\rm km\,s}^{-1}\else \,km\,s$^{-1}$\fi}   
  
\def\farcs{\hbox{$.\!\!^{\prime\prime}$}}  
  
\def\arcmin{\hbox{$^\prime$}}  
\def\arcsec{\hbox{$^{\prime\prime}$}}  
\def\secd#1.#2{ #1\farcs#2 }               
  
\def\mincir{\ \raise-2.truept\hbox{\rlap{\hbox{$\sim$}}\raise5.truept  
    \hbox{$<$}\ }}  
\def\magcir{\ \raise-2.truept\hbox{\rlap{\hbox{$\sim$}}\raise5.truept  
    \hbox{$>$}\ }}

\def\nii{[N~{\sc ii}]}  
\def\oiii{[O~{\sc iii}]}  
\def\ha{H$\alpha$}  
\def\stry{{Str\"omgren {\it y}}}  
\def\sexb{\object{Sextans~B}}

\def\hii{H~{\sc ii}}  
\def\heii{He~{\sc ii}}  
\def\sii{[S~{\sc ii}]}  
  
\documentclass{aa}  
\usepackage[dvips]{graphics}  
\begin{document}  
\title{The Local Group Census: planetary nebulae in   
Sextans~B\thanks{Based on observations obtained at the 2.5m~INT telescope  
operated on the island of La Palma by the Isaac Newton Group in the Spanish  
Observatorio del Roque de Los Muchachos of the Instituto de Astrofisica de  
Canarias.}}  
 
\author{ 
L. Magrini     \inst{1},   
R.L.M. Corradi \inst{2},  
N.A. Walton    \inst{2},  
A.A. Zijlstra  \inst{3}
D.L. Pollacco  \inst{4},\\  
J.R. Walsh     \inst{5},  
M.   Perinotto \inst{1},  
D.J. Lennon    \inst{2},  
R.   Greimel   \inst{2}
}  
\offprints{R. Corradi\\   
e-mail: rcorradi@ing.iac.es}  
\institute{ 
Dipartimento di Astronomia e Scienza dello Spazio, Universit\'a di 
Firenze, L.go E. Fermi 2, 50125 Firenze, Italy 
\and   
Isaac Newton Group of Telescopes, Apartado de Correos 321, 38700 Santa   
Cruz de La Palma, Canarias, Spain  
\and  
Physics Department, UMIST, P.O. Box 88, Manchester M60 1QD, UK 
\and 
School of Pure and Applied Physics, Queen's University Belfast, Belfast BT7 
9NN,
Northern Ireland, UK
\and   
ST-ECF, 
E.S.O., Karl-Schwarzschild-Strasse 2, 85748 Garching bei M\"nchen, Germany 
} 
 
\date{received date; accepted date}  
 
\abstract{   
Five planetary nebulae (PNe) have been discovered in the nearby dwarf
irregular galaxy.  Emission line images were obtained using the Wide
Field Camera of the 2.5m Isaac Newton Telescope (INT) at La Palma (Spain). The
candidate PNe were identified by their point-like appearance and
relatively strong \oiii\ emission-line fluxes. They are located within
a galactocentric distance of 2.8 arcmin, corresponding to 1.1~kpc at
the distance of \sexb. Luminosities are in the range
1800--5600$\,L_\odot$. \sexb\ is one of the smallest dwarf irregular
galaxies with a PN population. The number of PNe detected suggest an
enhanced star formation rate between 1 and 5 Gyr ago.
}
\authorrunning{Magrini et al.}   
\titlerunning{Planetary nebulae in Sextans B}    
\maketitle

\section{Introduction}   
The galaxies of the Local Group (LG) are our closest neighbors in the
Universe.  All the galaxies within 1.25 Mpc from its barycentre are
considered members of the LG (Courteau \& van den Bergh \cite{courteau}).
The LG contains galaxies covering a large range of luminosities,
metallicities, and morphological types, but the large majority are
dwarf irregular and spheroidal galaxies. Their close distance enables
the study of their content in much more detail than in more distant
galaxies.  For this reason, a new deep survey, the Local Group Census
(LGC\footnote{See {\tt http://www.ing.iac.es/~rcorradi/LGC}})
has been undertaken in all the LG galaxies above Dec=$-30^\circ$. The
survey aims at highlighting all populations with strong emission
lines, such as, H~II regions, planetary nebulae (PNe), supernova remnants,
luminous blue variables, and Wolf-Rayet stars.
  
One of the first targets of the LGC was the dwarf irregular galaxy \sexb, which
at a distance of 1.32$\pm$0.12~Mpc (Sakai et al. \cite{sakai}) is located at
the outer fringes of the Local Group. Although its radial velocity is
consistent with membership of the LG, it might form, together with
\object{NGC~3109}, \object{Antlia}, \object{Sextans~A} and perhaps
\object{AM1013-394A}, a physical association of galaxies lying just beyond the
boundaries of the LG, which would be the nearest association outside the LG
(van den Bergh \cite{vdB}).  The resolved stellar population of \sexb\ is
dominated by red giants (Sakai et al. \cite{sakai}), with a prominent
contribution of intermediate-age asymptotic branch stars. A young population is
also present, indicating a complex, low-rate and likely discontinuous star
formation throughout the galaxy lifetime (Tosi et al. \cite{tosi}; van den
Bergh \cite{vdB}).  A dozen \hii\ regions are known (Strobel et al.
\cite{strobel}), but no PNe have been identified prior to this study (Jacoby \&
Lesser \cite{jacoby81}).
  
The narrow-band images (\oiii, \ha + \nii, \sii, \heii) presented in 
this paper show five new candidate PNe in 
\sexb. The observations are described in Sect.~2, and data reduction and 
analysis in Sect.~3.  In Sect.~4, we present the candidate PNe, and 
discuss the results in Sect.~5. 
   
\section{Observations}  
  
We observed \sexb\ (\object{DDO~70}, 10h\,00m\,00s +05d\,19m\,56s, J2000.0)
using the prime focus wide field camera (WFC) of the 2.54~m Isaac
Newton Telescope (La Palma, Spain), on February and April 2001.  The
panoramic detector of the WFC consists of four thinned EEV CCDs of
2048$\times$4096 pixels each, and with a pixel  scale of 0.$\arcsec$33.
The $34\arcmin \times34\arcmin$ field of view of the camera covered
the entire galaxy.  Four narrow-band filters were used, with the
following central wavelengths and full widths at half maximum (FWHMs):
\oiii\ (500.8/10.0~nm), \ha + \nii\ (656.8/9.5), \heii\  
(468.6/10.0) and \sii\ (672.5/8.0).  A \stry\ filter, centred at
555.0~nm and with a FWHM of 30.0~nm, was also used to obtain off-band
images for continuum subtraction.
  
Exposure times were 600 s for {\it y}, 1200 s for \oiii\ and \ha +  
\nii, and 3600 s for \heii\, and \sii\, (split into three  
sub-exposures).  The seeing was 1\arcsec\, through the \oiii, \ha + \nii,  
\sii\ and {\it y} filters, and 1\arcsec.5 through the \heii\ filter.  
Several observations were made each night of the spectrophotometric  
standard stars: BD+33~2642, G191-B2B and Feige~34 (Oke \cite{oke}).  
  
\section{Data reduction and analysis}   
  
The frames were processed using IRAF. They were de-biased and flat-fielded
using the ING WFC data-reduction pipeline (Irwin \& Lewis \cite{irwin}).
Subsequently, they were corrected for geometrical distortions and aligned to
the \oiii\ frames.  The sky background was determined in  areas far from the
galaxy and then subtracted from all images. 
  
Emission-line objects were identified using a standard on-band/off-band
technique (Magrini et al. \cite{magrini00}). Briefly, the stellar continuum of
\sexb\ (\stry\ images, properly scaled) was subtracted from the narrow-band
images in order to highlight emission-line objects.  Aperture photometry was
applied using the IRAF task APPHOT in the continuum-subtracted frames.
Photometric errors were derived from photon statistics for both sources and
background with APPHOT.  They are of few per cents both for \oiii\ and for
\ha+\nii\ fluxes. 

The instrumental magnitudes were calibrated by convolving the spectrum of the
standards used (Oke \cite{oke}) with the response curve of each filters.  An
interstellar extinction of E(B-V)=0.02 was used (de Vaucouleurs et al.
\cite{deV}), following the Seaton (\cite{seaton}) prescription.  The fluxes in
the \oiii\ line at $\lambda$=500.7~nm were corrected for the contribution of
the companion oxygen line at $\lambda$ 495.9~nm, which amounts to 20\%\ of the
total flux in the relatively broad filter used.  Including the uncertainties in
all steps, we estimate that the total absolute error on the absolute fluxes
quoted in this paper is $\le$20$\%$.

The astrometric solution for the \oiii\ frames was also obtained through the
WFC pipeline using the USNO A2.0 catalogue (Monet et al. \cite{monet}).  The
solution has an {\it r.m.s.} accuracy of 0$''$.3.

\begin{table*}   
\caption{PN candidates in \sexb. Positions and reddening corrected    
\ha+\nii\ and \oiii\ emission-line fluxes (in units of  
10$^{-16}$~erg~cm$^{-2}$~s$^{-1}$) are given, and luminosity in
solar units.}  
\begin{center}
\begin{tabular}{l r r r r r l}    
\hline   
Identification & \multicolumn{2}{c}{R.A. (2000.0) Dec.} & $F_{\rm [OIII]}$ & 
$F_{\rm H\alpha}$ & m$_{\rm [O~III]}$  & $L$ \\ 
\hline   
\object{SexB PN1} & 9 59 53.06  & 5 18 52.1  & 16.0 & 7.0  & 23.24 & 2000 \\    
\object{SexB PN2} & 9 59 56.50  & 5 19 29.5  & 11.2 & 8.1  & 23.64 & 2300 \\  
\object{SexB PN3} & 9 59 56.64  & 5 19 52.6  & 19.2 & 13.0 & 23.05 & 3600 \\  
\object{SexB PN4} & 10 00 00.19 & 5 20 37.2  & 12.8  & 6.6  & 23.49 & 1800 \\ 
\object{SexB PN5} & 10 00 10.48 & 5 20 31.9  & 38.4  & 20.0 & 22.30 & 5600 \\ 
\hline
\end{tabular}   
\end{center}
\end{table*}

\section{Candidate planetary nebulae}  
 
The \ha+\nii\ and \oiii\ continuum-subtracted images (Fig.~1, left
panels) show a number of emission-line objects. Some of them are
clearly extended \hii\ regions, which were previously studied by Moles
et al. (\cite{moles}), Strobel et al. (\cite{strobel}), Hunter et al.
(\cite{hunter93}), Hunter \& Hoffman (\cite{hunter99}), and Roye \& Hunter
(\cite{roye}). We note that these regions
form an approximate ring, with another H$\alpha$ nebulosity in the
centre of the ring, perhaps close to the nucleus.

Candidate PNe in \sexb\ were selected using the following criteria 
(Magrini et al. \cite{magrini00}, \cite{magrini01}): 
  
i) they should appear both in the \oiii\ and \ha+\nii\ images but not 
in the continuum frame; 
  
ii) they should have a stellar point spread function: at the  
distance of \sexb. PNe are normally 0.1--1 pc in diameter,
corresponding to 15--150 mas at the distance of \sexb.
  
Five objects in \sexb\ fulfill the criteria above; these new
candidate PNe are listed in Table~1 and marked in Fig.~1.  Only
\oiii\ and \ha+\nii\ fluxes are quoted in Table~1: they were not
detected in \heii\ and \sii.  The upper limit to these latter fluxes
is estimated to be 10$^{-16}$~erg~cm$^{-2}$~s$^{-1}$. The \oiii\
fluxes were converted into equivalent V-band magnitudes following
Jacoby (\cite{jacoby89}):
\begin{equation}  
m_{\rm [O~{III}]}=-2.5\log F_{\rm [O~{III}]} -13.74.  
\nonumber  
\end{equation}  

The luminosity was obtained from the relation $L \approx 150 \times L(H\beta)$
(Zijlstra \&\ Pottasch \cite{zijlstra89}),which is correct when the nebula is optically thick. 
This assumes that the [N~II] lines
make a negligible contribution to the H$\alpha$ line, which is likely correct
at this low metallicity. The luminosities are close to but below the value of
$7000\,\rm L_\odot$, as expected for progenitor masses $< 2\,\rm M_\odot$.

This is the first identification of PNe in \sexb: previous surveys did not find
any suitable candidate (Jacoby \& Lesser \cite{jacoby81}).  The five candidate
PNe are distributed over an area of 1\arcmin.5$\times$4\arcmin.5, corresponding
to a linear, projected size of 0.6~kpc$\times$1.7~kpc.  Three of them are found
in regions densely populated by stars, whilst the other two (\object{SexB PN1}
and \object{SexB PN5}) are located in the outskirts of \sexb, at projected
distances from the galaxy centre of 0.8 and 1.1~kpc. All the candidate PNe have
\oiii/(\ha+\nii) flux ratios between 1.4 and 2.3. These are typical values for
Galactic PNe ({\it cf.} Magrini et al. \cite{magrini00}), although the oxygen
abundance of \sexb\ is only 0.16 times solar (Moles et al. \cite{moles}).
  
\subsection{Completeness limits} 
 
The detection of the five PNe raises the question: how many more remain to be
discovered? Some PNe could have been missed in the most crowded region of
\sexb.  We have estimated this number by adding `artificial stars' within the
range of luminosities expected for PNe (Jacoby \cite{jacoby89}) in both the
\oiii\ and the off-band, {\it y}, frames using the IRAF task ADDSTAR. These
artificial stars were recovered following the same procedures used to detect
PNe. A 3$\sigma$ detection limit was adopted in the off-band frame.  The
recovery rate is significantly lower in the continuum-subtracted frames
because of the larger noise produced in the scaling and subtraction of two
images. This effect is dominant in recovering emission-line objects at faint
magnitudes.

The incompleteness is a combination of the probability of missing an
object in the emission-line image and the probability of wrongly
identifying a star in the continuum frame. Incompleteness is defined
as a recovery rate of artificial stars less then 50\% ({\it e.g.} Minniti \&
Zijlstra \cite{minniti}).  This was computed in different regions of \sexb,
and for a range of assumed magnitudes. We find that incompleteness
occurs for emission-line objects fainter than $m_{\rm
[O~{III}]}$=24.5, located within 1\arcmin.5 from the centre of the
galaxy. 

Counting all stars brighter than this completeness limit, we find that
the total luminosity of these stars in the inner regions is about 40\%
of the total.  This agrees (roughly) with 3 of the 5 PNe found in this
region.  As completeness is better than 50\%\ in this area, we
conclude that at most five PNe brighter than $m_{\rm [OIII]}$=24.5
may have been missed there.

\subsection{Total PNe population size}

The number of candidate PNe that we found in \sexb\ is consistent
with the expected population size for this galaxy, which can be
estimated from stellar population models. For a simple ({\it i.e.} coeval
and chemically homogeneous) stellar population, the number of stars
$n_j$ in any post-main-sequence phase $j$ (Renzini \& Buzzoni \cite{renzini}) is
given by
  
\begin{equation}  
n_{j}=\dot{\xi} L_T t_j,   
\nonumber  
\end{equation}  
  
where $\dot{\xi}$ is the specific evolutionary flux (number of stars
per unit luminosity leaving the main sequence each year), $L_T$ the
total luminosity of the galaxy, and $t_j$ the duration of the
evolutionary phase $j$ ($\le$20~000 yrs for the PN phase).  Adopting a
bolometric luminosity of \sexb\ of $\sim$10$^8$~$L_\odot$ (van den
Bergh \cite{vdB}), and a specific evolutionary flux between
$5\times10^{-12}$ and $2\times10^{-11}$~yr$^{-1}$~$L_\odot^{-1}$
(cf. Renzini \& Buzzoni \cite{renzini}, Mendez et al. \cite{mendez}), the
corresponding population size of PNe in \sexb\ is between 5 and 20 objects.
This appears to be in good agreement with the number of PNe discovered.
However, the [O~III] luminosity of a PN rapidly declines once the nebulae
becomes optically thin, and the phase during which a PN is bright enough to be
discovered here is much shorter, typically 5000 yr (assuming an expansion
velocity of 20 km/s and a radius of 0.1 pc).  Thus, the number of PNe may be a
little higher than expected.

Figure 2 shows the number of known PNe for all nearby galaxies, using
data in van den Bergh (\cite{vdB}), with three exception: M33 now has 131
known PNE (Magrini et al. \cite{magrini01}), whilst for the Sagittarius dwarf
spheroidal 3 PNe are now known (Walsh, priv. comm.), and for WLM both
published PNe have been shown to be normal stars (Minniti \&\ Zijlstra
\cite{minniti}).  For the Milky Way (MW), about 1100 PNe are now known, but the
total number has been estimated to be as high as 23000 (Zijlstra \&\
Pottasch \cite{zijlstra91}). Figure 2 plots the number of PNe versus the V-band
luminosity of the galaxy, in solar units.  Eq. (1) suggests that this
relation should be strictly linear, for a uniform star formation
history.  The dashed line gives the expected total number of PNe
scaling from 23000 in the MW. The dotted line is fitted to the known
number in the LMC, where surveys have been most complete.  It is clear
that only the bright tip of the PN luminosity function can be
detected. Figure 2 also shows that for galaxies with $M_V > -12.5$, one
would not expect to detect the PN population.

Given its distance, the PN census in \sexb\ may be expected to be less
deep than that in the LMC. However, the two fall on the same
line. This could be explained if \sexb\ has a large proportion of
intermediate-age stars, with a significant fraction of this star
formation over the past 5 Gyr or so, compared to the LMC:
a younger population has a higher death rate and will produce more PNe.
The star formation in the Universe has declined rapidly from its peak around $z
\approx 1.5$ (Lilly et al. \cite{lilly}; Madau et al. \cite{madau}); however,
\sexb\ may be an
example of a galaxy with a more delayed star formation history.  The
distribution in Fig. 2, once completeness has been reached, can be
used to measure the star formation history for intermediate-age stars.

The small population size prevents us from building a meaningful 
Planetary Nebulae Luminosity Function (PNLF) for \sexb, a property 
that is generally used as an extragalactic distance indicator (Jacoby 
\cite{jacoby89}). In fact, the absolute magnitude of the bright cutoff of the 
PNLF depends on the size of the PN population. For a small population 
as in \sexb, the bright PNe defining the ``universal'' cutoff of 
the PNLF ($m_{\rm [O~{III}]}^\star=-4.48$, Jacoby 1989) are not 
observed owing to the poor sampling.  The simulations of Mendez et 
al. (\cite{mendez}) show, that even for a population size of 500~PNe, typical 
of the LMC and much larger than in \sexb, no PNe with absolute 
magnitude lower than $-4.0$ are expected.  Therefore, the absolute 
magnitude of the brightest PN in \sexb, equal to $-3.59$ mag, is 
qualitatively consistent with its small population size, and can only 
provide an upper limit of $\sim$2~Mpc for the distance of the galaxy. 
 
Figure 2 and the completeness analysis above suggest that the PN census
in \sexb\ is now fairly complete.  In many other galaxies (especially
IC~10), significant PN populations remain to be found.

\section{Summary and conclusions}  
 
Five new candidate PNe have been discovered in the dwarf irregular 
galaxy \sexb. 
This is a notable result considering the limited number of PNe known in the
other dwarf galaxies of the LG. In fact, a total of four PNe are known in the
dwarf spheroidal companions of the Milky Way: three PNe in Sagittarius (Walsh
et al. \cite{walsh97}; Dudziak et al. \cite{dudziak}), and one in Fornax
(Danziger et al. \cite{danziger}). Another few PNe have been discovered in the
dwarf companions of M~31, and one in NGC~6822 (Ciardullo et al.
\cite{ciardullo}; Jacoby \& Ford \cite{jacoby86}). 
  
The number of candidate PNe detected in \sexb\ is in good agreement 
with the PNe population size expected from stellar evolution models.

\begin{figure*} 
\resizebox{\hsize}{!}{\includegraphics{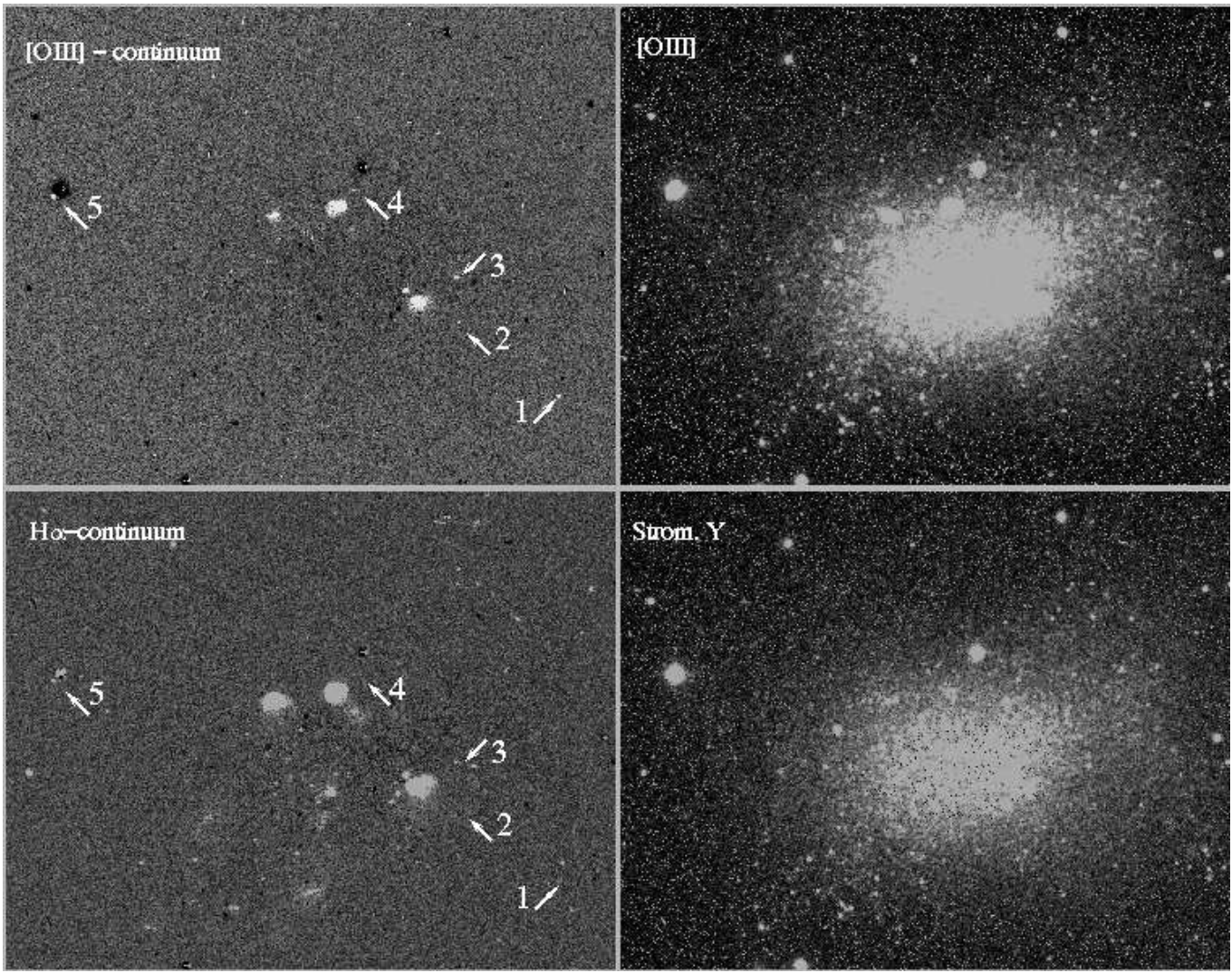}} 
\caption{INT+WFC images of Sextans~B. 
The size of each frame is 5\arcmin.1$\times$4\arcmin.0, {\it i.e.} a small 
fraction of the whole field of view of the WFC.  North is at the top, 
East to the left.  In the continuum subtracted images, candidate PNe 
are indicated by the arrows and marked with the identification number  
as given in Table~1.} 
\end{figure*}

\begin{figure*} 
\resizebox{\hsize}{!}{\includegraphics{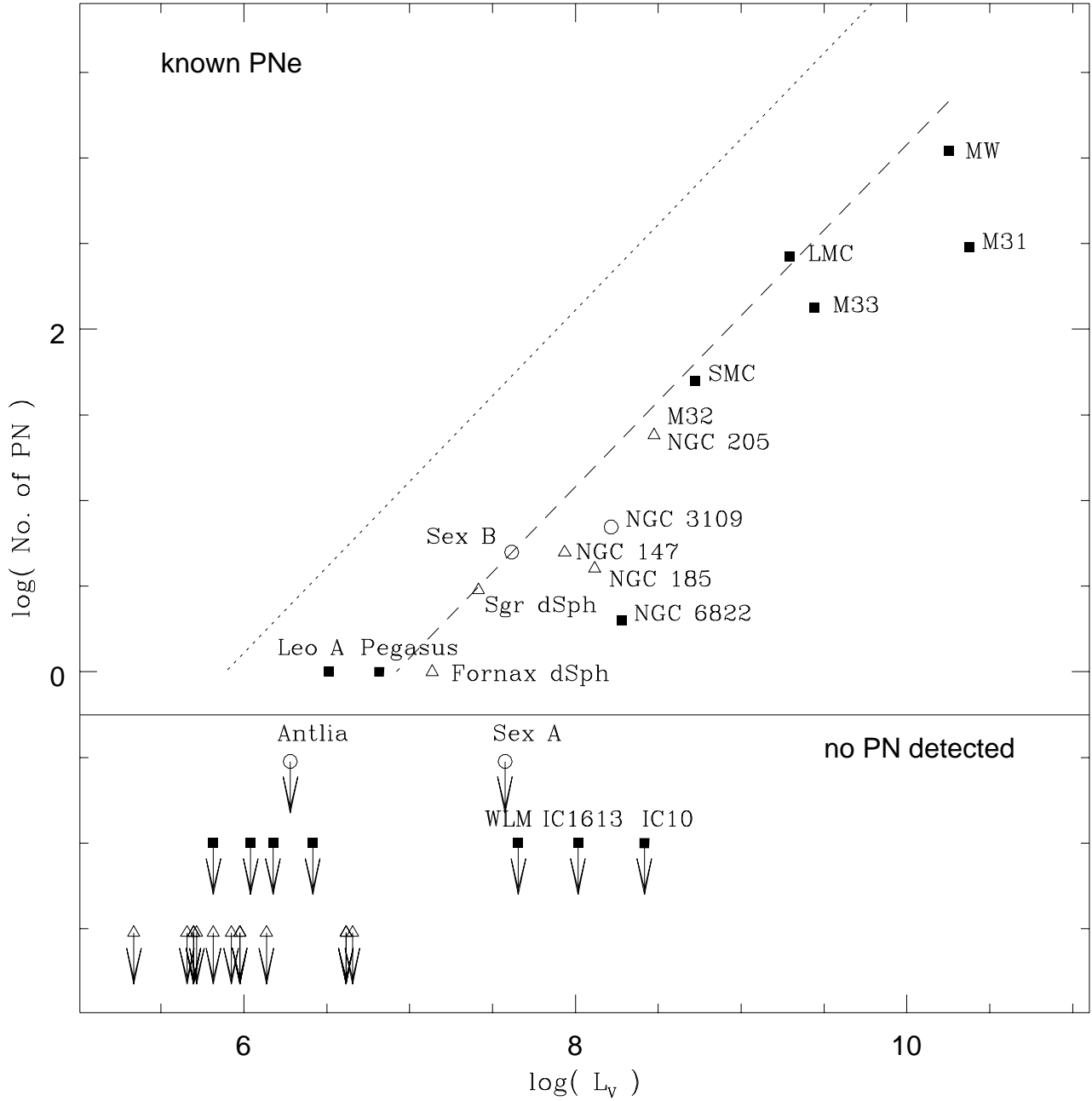}} 
\caption{The number of PNe of galaxies in or near the Local Group,
versus the V-band luminosity in solar units. The dotted line shows
the expected numbers based on a total number of PNe in the MW of
23000. The dashed line is fitted to the known population in the LMC.
Filled squares indicate LG gaseous galaxies, triangles indicate spheroidal
galaxies and open circles show the NGC 3109 group.
} 
\end{figure*} 
  
\end{document}